# Relayed-QKD and switched-QKD networks performance comparison considering physical layer QKD limitations

N. Makris, A. Papageorgopoulos, P. Konteli, I. Tsoni, K. Christodoulopoulos, G. T. Kanellos\*, D. Syvridis

*National and Kapodistrian University of Athens, Athens, Greece*
*\*gtkanellos@di.uoa.gr*

**Abstract:** We experimentally evaluate the SKR generation for unoptimized QKD pairs in switched QKD and compare the performance of the switched-QKD with relayed-QKD networks to reveal they perform better for short distances and at large networks. © 2023 The Authors

1. **Introduction**

Prepare and Measurement Quantum key distribution (PM-QKD) is the most mature scheme for QKD and commercial Discrete Variable (DV) QKD systems have already been deployed in real networks [1]. PM-QKD rely on point-to-point (p2p) deployments and thus QKD networks currently require a relay mechanism that takes place in intermediate trusted nodes to propagate the QKD keys across the network while several standards for relayed-QKD have been recently released to regulate this operation [2]. However, in common relayed-QKD networks with N nodes, each equipped with one Alice and one Bob QKD (thus N pairs in total), the nodes need to share these resources not only to communicate with the adjacent nodes but also to serve as intermediary to relay keys for non-neighbouring nodes, significantly limiting the resource usage efficiency particularly in long links with multiple intermediate nodes (Fig. 1a). To scale QKD networks from a few p2p links to hundreds of seamlessly interconnected nodes will require real dynamic network reconfiguration capabilities in the physical layer to allow for optimal hardware resource allocation and usage, much as the classical optical networks example had to move from the $1^{st}$ Gen. p2p WDM links to $3^{rd}$ Gen. OTN-Optical transport networks with the adoption of reconfigurable optical add-drop multiplexers (ROADM). To this end, several dynamic or switched QKD network implementations [3] have been recently demonstrated employing low loss optical switches (LLOS) in front of an Alice and/or Bob QKD devices to interface those with any other dual device in the network (Fig. 1b) allowing for a direct QKD link between any node in the network and thus removing the need to unnecessarily consume keys in chained QKD links.

Though, this proves a significant scaling advantage for the switched-QKD, its overall network performance in terms of total secret key capacity may be significantly reduced due to the physical impairments of the QKD technology. In particular, the QBER and Secret Key Rate (SKR) in every QKD link of the switched-QKD network are affected by 1) the additional losses of the optical switches 2) the exponential decay with the link distance, thus providing a significantly lower SKR for the longer distance diagonal QKD links of the network. In addition, switched–QKD comes with the requirement that every Alice needs to be able to communicate with any other Bob in the network. Although this might be a true assumption in most cases, arbitrary QKD pairs are not identical in hardware and would require fine tuning of their operational parameters in terms of wavelength emission and filtering, phase/differential delays/polarization bases matching and other timing/synchronization or software parameters to achieve optimal performance and this would still not guarantee a uniform performance across all pairs. Thus, it is fair to assume that any unmatched arbitrary QKD pair in a switched-QKD network is to some extent unoptimized and would exhibit a performance variation from its matched counterpart. Finally, on a switched-QKD network each link is established for a fraction of the time and the key generation capability per link for any device is divided to N/2 (assuming that Alice and Bob each serves half). Moreover, to achieve a fair key generation, one would allocate a time-portion that is inverse to the SKR generation rate of the QKD link, spending, thus, (exponentially) more time in a link with distant nodes.

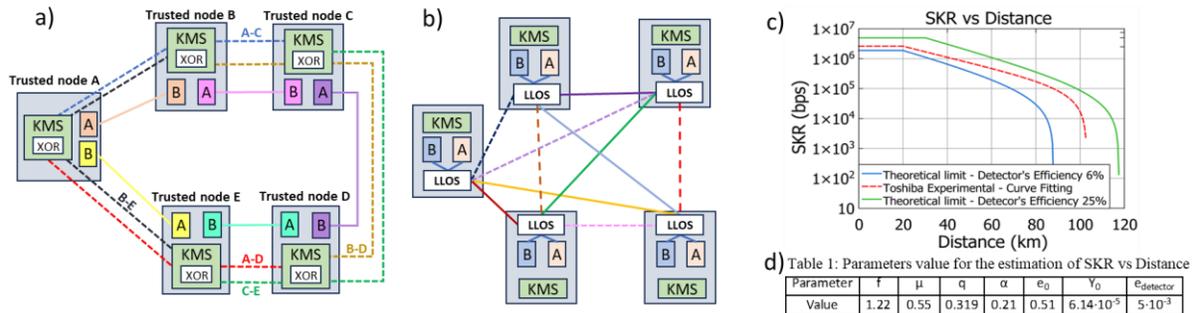

Fig. 1. a) Relayed-QKD network with 5-QKD pairs and multi-hop QKD links for non-neighboring nodes. b) Switched-QKD with 5 direct QKD links. c) Experimental and simulated SKR vs Distance for DV-QKD. d) Parameters used in the theoretical estimation of the SKR graph.

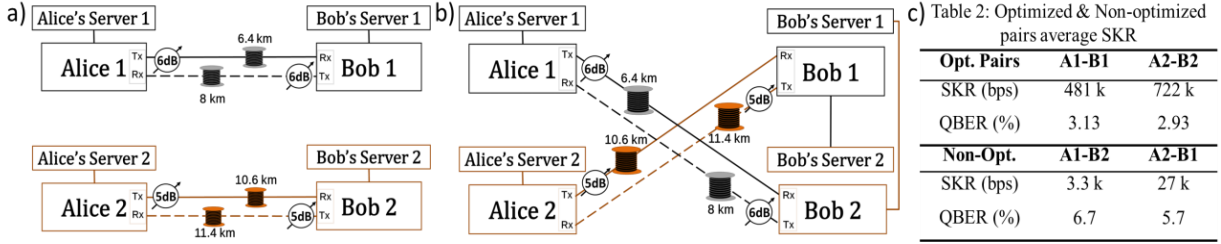

Fig. 2. a) Optimized QKD pairs and b) non-optimized QKD pairs. c) Average values for SKR and QBER for Opt. and non-opt. QKD pairs.

In this communication we aim to evaluate the performance of switched-QKD networks against relayed-QKD networks in terms of cumulative secret key capacity while maintaining fairness in SKR rates for each link. To this end, we first discuss the function F(a) of exponential decrease of SKR with distance/loss a. We then present experimental results for two unmatched pairs of commercial off-the-shelf QKD devices (Toshiba QKD4.2-MU/MB [7]) implementing an efficient BB84 protocol with decoy states and phase encoding [8] to reveal a significant ~20dB drop of SKR for unmatched pairs. Based on these findings and acknowledging that the SKR drop may significantly improve if the arbitrary pairs of the QKD systems are properly aligned, we evaluate the performance of the switched-QKD architecture for moderate penalty (5dB) in unmatched pairs. Finally, we compare with relayed QKD networks over a ring network topology, with equal length/attenuation links between adjacent nodes. We reveal that switched-QKD performs better for short distances and at large networks with high number of hops.

## 2. Experimental evaluation of SKR vs loss and vs unmatched pairs

Fig. 1c depicts the experimental SKR vs distance obtained with the Toshiba QKD systems as well as the numerical simulation with detector efficiencies $\eta_{Bob}$ = 6 % and 25 % respectively, while Table 1 summarizes the parameters used for the SKR and QBER equations based on [9]. These values are used to define $f(a)$ in the Section 3 simulation. For the evaluation of the switched operation, we consider the connectivity as shown in Fig. 2a,b. Alice 1 connects to its optimized pair Bob 1 (A1B1) via a 6.4 km link fitted with 6 dB of attenuation to emulate the optical switches loss and Alice 2 connects to its optimized pair Bob 2 (A2B2) via a 10.6 km link fitted with 5 dB of attenuation simulating the optical switches loss. In each pair, QKD channels are at the O-band copropagating with two service channels (1529 nm and 1530 nm), while a third service channel at 1528 nm counter propagates in a separate fiber. For the switched operation, Alice 1 is manually switched to a non-optimized pair Bob 2 (A1B2) as shown in Fig. 2b, with the same distances and total attenuation as in the A1-B1 case, and Alice 2 connection is manually switched to a non-optimized pair Bob 1 (A2B1) with the same distances and total attenuation as in the A2-B2 case. The results for up to 18h of operation are shown in detail in Fig. 3 where the total key generation for A1-B2 is 221 Megabits and for A2-B1 is 1.8 Gigabits. Table 2 summarizes the average values obtained for SKR and QBER in all the cases and reveals a ~14 dB drop for A2-B1 and more than 20 dB for A1-B2. However, we note that the QKD phase encoding protocol requires perfect matching between the differential interferometers (DI) used in Alice to temporally separate the weak coherent pulses and to properly recombine them in Bob. Imperfect matching will lead to poor visibility for the quantum states and therefore lead to increased QBER and reduced SKR. Such adjustments of the DI are normally performed by the servers driving the Alice and Bob units during initialization, but in our case, we only switched the optical QKD engines and maintained the original servers' connectivity (Server Bob 2 still drives Bob 1 and Server Bob 1 drives Bob 2), thus preventing the unmatched pairs recalibration.

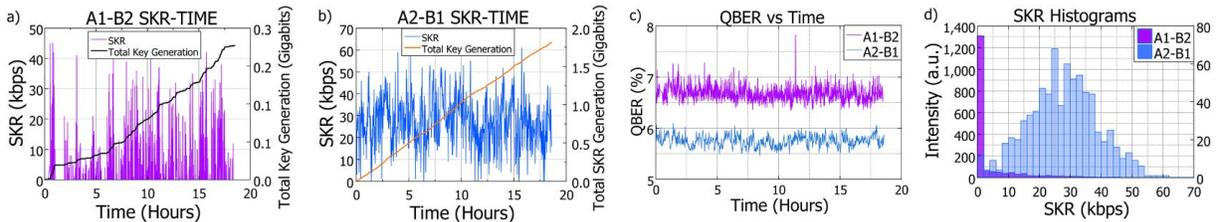

Fig. 3. a) A1-B2 SKR b) A2-B1 SKR c) QBER for switched QKD pairs d) SKR Histogram for switched QKD pairs.

## 3. Switched vs relayed QKD performance.

We compare the performance of switched to relayed QKD architecture over a ring topology, commonly used in metro, with equal-length/attenuation links while the SKR is given by function $f(a)$ as discussed in Section 2. A penalty of 5 dB was assumed for switched QKD to account for non-matching QKD pairs and losses in optical switches. We assume an all-to-all key consumption pattern from Security Application Entity (SAE) pairs. Our goal is to fairly equalize the SKR for all SAE pairs and maximize it to enhance the overall network security.

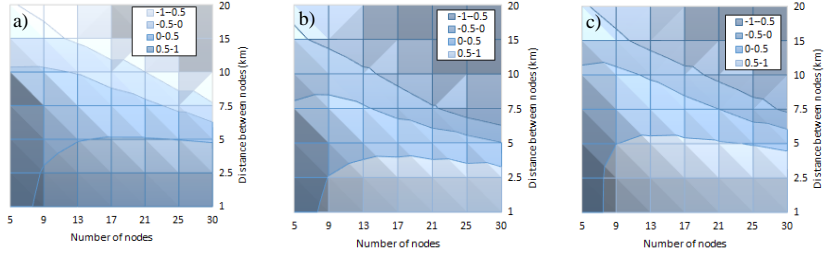

Fig. 4. Switched to relayed QKD $(G_S-G_R)/\max(G_S,G_R)$ normalized SKR difference for a) experimental SKR, b) low SKR and, c) high SKR.

In relayed QKD, Alice in node $n_i$ connects to its matched Bob in the next node $n_{i+1}$ and so on, to form a full ring. Nodes act as trusted relays to create keys for non-adjacent node pairs. In the ring topology, each QKD link feeds a total of $N^2-1/8$ node pairs (for $N$ odd) that have that link as an intermediary in their key generation path. Thus, in a ring with equal link attenuation ($A_e$) targeting a fair (equal) SKR, each SAE pair obtains a SKR equal to $G_R=8\cdot f(A_e)/(N^2-1)$. In switched QKD, we assume direct fibers from a node to any other node, interfaced through an optical switch to the Alice and Bod in the node (Fig. 1b). The lengths of all diagonal links are geometrically estimated based on a circle with diameter equal to the ring. A scheduling algorithm defines the configuration of the switches and formation of QKD links; in a ring with equal link lengths, scheduling is straightforward. To achieve fair SKR for node $n_i$, we configure it to communicate with half network nodes using its Alice and half with its Bob for a timeslot that is inversely proportional to the link SKR. In particular, $n_i$ would connect to $n_j$ for a portion $T_{ij}=2/f(A_{ij})$ of a total of $T_i=\Sigma_j T_{ij}$ time, where $A_{ij}$ is the attenuation of QKD link $(n_i,n_j)$. Then, each SAE pair obtains an SKR of $G_S=1/T_i$. This schedule is applied simultaneously across all nodes; thus, achieving an SKR of $G_s$ for all SAE pairs.

We compare the two architectures in terms of the maximum common key SKR. Relayed QKD divides the SKR of a QKD pair into $(N^2-1)/8$, while switched QKD divides it into $N/2$. However, switched QKD spends more time in long QKD links to compensate for the (exponentially) reduced SKR. We assume rings with varying number of nodes ($N$=5 to 30), and varying adjacent nodes distances (1 to 20 km) and attenuation coefficient of 0.21 dB/km. We examine three SKR functions $f(a)$ as calculated in Section 2 to emulate the different QKD performance in terms of SKR vs losses (distance). We plot the normalized SKR difference of switched to relayed QKD, i.e., $R=(G_S-G_R)/\max(G_S,G_R)$, ranging from -1 (relayed outperforms switched) to +1 (opposite). Fig. 4a plots R when using $f(a)$ for the experimentally fitted SKR function, showing that for distances less than 5 km, even with 30 nodes, switched QKD outperforms relayed QKD. For 7.5 km links, switched is better for up to 20 nodes, and for 10 km, it is better for up to 10 nodes. Short distances and a high number of hops favor switched QKD, as it remains in the flat area of the key generation function and/or there are no high attenuation links. For low $f(a)$ SKR performance Fig. 3b reveals a small reduction in the efficiency and applicability of the switched QKD architecture as opposed to the relayed network, while for high SKR (green $f(a)$ curve of Fig. 1c), Fig. 3c reveals that switched QKD is better for 20 nodes/10 km links.

4. Conclusion

We presented experimental results for two unmatched pairs of commercial phase-encoding QKD devices revealing a significant drop of SKR for unmatched pairs. We also evaluated the performance of the switched-QKD and compared with relayed QKD networks over a ring network topology and revealed that switched-QKD performs better for short distances and at large networks with high number of hops.

5. Acknowledgements

This work was funded by the EU project QSNP (GA 101114043) and the HellasQCI project (GA 101091504).